\documentclass[pdflatex,sn-mathphys-num]{sn-jnl}

\usepackage{graphicx}
\usepackage{longtable}
\usepackage{multirow}
\usepackage{amsmath,amssymb,amsfonts}%
\usepackage{amsthm}%
\usepackage{mathrsfs}%
\usepackage[title]{appendix}%
\usepackage{xcolor}%
\usepackage{textcomp}%
\usepackage{manyfoot}%
\usepackage{booktabs}%
\usepackage{algorithm}%
\usepackage{algorithmicx}%
\usepackage{algpseudocode}%
\usepackage{listings}%

\theoremstyle{thmstyleone}%
\theoremstyle{thmstyletwo}%

\theoremstyle{thmstylethree}%

\begin{document}

\title[Article Title]{Scalar tidal response of static and rotating black holes in anti-de Sitter spacetime}


\author[1,2]{\fnm{Yusmantoro}}\email{yusmantoroyusman@gmail.com}

\author[1,2]{\fnm{Freddy} \sur{Permana} \sur{Zen}}\email{fpzen@fi.itb.ac.id}
  
\author[3,2]{\fnm{Hadyan} \sur{Luthfan} \sur{Prihadi}}\email{hadyanluthfanp9@gmail.com}

\affil[1]{\orgdiv{Theoretical High Energy Physics Group}, \orgname{Department of Physics, FMIPA, Institut Teknologi Bandung}, \orgaddress{\street{Jl. Ganesha 10}, \city{Bandung}, \country{Indonesia}}}

\affil[2]{\orgdiv{Indonesia Center for Theoretical and Mathematical Physics (ICTMP)}, \orgname{Institut Teknologi Bandung}, \orgaddress{\street{Jl. Ganesha 10}, \city{Bandung}, \postcode{40132}, \country{Indonesia}}}

\affil[3]{\orgdiv{Research Center for Quantum Physics}, \orgname{National Research and Innovation Agency (BRIN)}, \orgaddress{\city{South Tangerang}, \postcode{15314}, \country{Indonesia}}}


\abstract{In this work, we investigate the tidal response of both static and rotating black	holes in anti-de Sitter spacetime. We perform the neutral scalar field perturbation for Schwarzschild and Kerr black holes. On the other hand, charged scalar field are used for perturbing Reissner-Nordstrom and Kerr-Newman black holes. We find that the tidal Love number for all black holes are always nonvanishing due to the presence of cosmological constant. In contrast, tidal dissipation can vanish for certain conditions depending on the scalar field's frequency and the black hole's parameters.}




\maketitle

\section{Introduction}\label{sec1}

The direct observation of gravitational waves (GWs) by LIGO in 2015 has opened a new way to test and improve Einstein's general theory of relativity especially in strong gravitational field \citep{abbott2016binary,abbott2020prospects,cahillane2022review}. GWs observation has attracted numerous experimental and theoretical physicists because it carries information describing and probing the properties of astrophysical compact objects \citep{baibhav2021probing,vijaykumar2023probing}. One of the most significant advantage of GWs observation is to provide a powerfull tool for investigating the nature of black holes. It is supported by the new evidence that GWs emitted by binary black hole merger has been observedn \citep{abbott2016observation,abbott2023gwtc}.

Another quantity that is strongly related to black hole's property is the tidal deformability or tidal response. It shows response of black hole's multipole moments to external tidal perturbation \citep{nair2024asymptotically}. According to linear response theory, tidal response is a linear combination of real and imaginary parts namely tidal Love numbers (TLNs) and tidal dissipation numbers (TDs), respectively\citep{chia2021tidal,hinderer2008tidal,flanagan2008constraining,damour2009relativistic}. TLNs is a conservative quantity encoding black hole's deformation because of its interaction with the tidal environment. While TDs identifies the lost of energy during interaction because of the viscocity around black hole's spacetime \citep{poisson2014gravity}.

In general relativity formulation, previous researches have shown that TLNs of static and slowly rotating asymtotically flat black holes are zero  \citep{bhatt2023addressing,bhatt2025rotating,binnington2009relativistic,le2021tidal,chakrabarti2013new,landry2015tidal,charalambous2021vanishing,hui2021static,gounis2025vanishing}. From this, according to Einstein's theory we can conclude that black holes are capable of keeping their gravitational dominance among astrophysical objects in the universe.
Some studies have demonstrated that zero value of TLNs can be explained by symmetry argument known as ladder symmetry or love symmetry \citep{hui2022ladder,charalambous2022love,sharma2024exploring,rai2024ladder}. However, it should be noted that the existence of love symmetry does not guarantee the vanishing of TLNs \citep{charalambous2022love}. It is because the presence of mass changes the eigenvalue of Casimir operator leading to regular solution of static scalar field at the horizon \citep{charalambous2022love}. 

TLNs of asymtotically flat black holes due to massless scalar field perturbation does not always vanish. In ref \cite{ma2025charging},  
the charged massless scalar field is considered for perturbing charged black hole. TLNs is non-vanishing and is inversely proportional to the scalar field's charge in its infinitesimal limit.Hence, discontinuity of TLNs between neutral and charge emerge as a nontrivial problem despite the ability of charge quantization mechanism for removing the divegence.   

Frequency-dependent of the static TLNs of Kerr black hole has been computed from scattering amplitude in effective field theory \citep{saketh2024dynamical}. Explicit results of TLNs in all order of spin can be expressed by taking low frequency approximation. In this formalism, tidal response is strongly related to local worldline couplings consisting of curvature time derivative, spin-curvature time derivative, and quadrupole-octupole mixing operators. Additionally, the known previous results of TLNs can be reproduced for linear and nonlinear order.

The calculation of TLNs in the lower dimensional black holes has also been obtained. Previous study has revealed that TLNs of $(2+1)$ dimensional rotating BTZ black hole is found to be nonzero and has scale-depending behaviour \citep{bhatt2024scalar}. However, TLNs vanishes for axisymmetric scalar perturbation for extremal and non-extremal cases. TLNs have been calculated as well for charged and rotating BTZ black hole by employing slowly varying scalar field approximation in the near horizon. The results are similar to that of uncharged one. 

The study of TLNs can be generalized to higher dimensional black holes \citep{rodriguez2023love}. 
TLNs of Reissner-Nordstrom and Schwarzschild black holes in $D$ dimensions are generally nonzero. Unfortunately, their exact values are difficult to obtain but vanish for specific values of $D$, for instance $D=4$ \citep{ma2025charging}. Despite nonvanishing TLNs of higher dimensonal black holes, they still possess Love symmetry. But, its eigen value does not belong to highest weight $SL(2,R)$ representation yielding non-vanishing TLNs.

The dynamics of scalar field in the near horizon of non-extremal black holes is also essential to study in Kerr/CFT correspondence \citep{guica2009kerr,bredberg2011lectures,compere2017kerr,sakti2019kerr,sakti2020kerr,sakti2022rotating}. The separable Klein-Gordon equation of the corresponding field can be reproduced by quadratic Casimir operator formed by generators of $SL(2,R)$ algebra. The emergence of this symmetry is called hidden conformal symmetry \citep{sakti2019deformed,sakti2020hidden,sakti2023hidden}. Unfortunately, it has been known that such generators are only well-defined for Schwarzschild black hole especially under $\phi\rightarrow\phi+2\pi$ identification \citep{bertini2012conformal}. On the other hand, Love symmetry is found to be globally well-defined and capable of replacing generators for hidden conformal symmetry in spherically and non-spherically symmetric black holes in four and higher dimensions. Consequently, love symmetry and hidden conformal symmetry are related to each other \citep{gray2025love}.

Nonvanishing TLNs for four-dimensional black holes can also be obtained by adding positive or negative cosmological constants to Einstein field equation resulting black holes in de-Sitter(dS) or anti-de Sitter(AdS) spacetimes, respectively \citep{nair2024asymptotically,franzin2025tidal}. This study is relevant because dS spacetime accommodate the explanation of expanding universe where cosmological constant plays an important role as the dark energy. On the other hand, physicists pay attention AdS spacetime due to its correspondence to conformal field theory in lower dimension. This is a conjectured relationship which is famously known as AdS/CFT correspondence \citep{maldacena1999large,petersen1999introduction}.

Kerr/CFT correspondence is regarded as the extension of AdS/CFT correspondence for Kerr black hole. Scattering near horizon and black hole's entropy can be recalculated using conformal field theory. This conjecture shows quantum signature of an object having strong gravitational field. Therefore, it provides a new and significant insight to the discovery of quantum gravity.

One intriguing results regarding quantum gravity and TLNs is that there is a possibility of TLNs as the signature of loop quantum gravity \citep{motaharfar2025loop}.
Quantum gravity effect emerges as the alternative solution to resolve the singularity of black hole. It has been revealed that quantum black hole possess a nonzero TLNs which is opposite to its classical counterpart. Additionally, using spin $0,1,$ and $2$ perturbations, TLNs is found to have negative values where its magnitude increases as the black hole's mass decreases. 

Based on above description, we argue that the study of TLNs in AdS black holes is essential and provide a positive impact on the signature of quantum gravity.
TLNs of Schwarzschild AdS black hole has been investigated using Regge-Wheeler gauge and Kodama-Ishibashi formalism \citep{franzin2025tidal}. In this spacetime, ladder symmetry does not exist yielding nonvanishing TLNs. Moreover, TLNs result has been compared to that of black branes in AdS and they are in agreement in the certain regime.
Ref \citep{nair2024asymptotically} find that TLNs of asymtotically de Sitter compact objects can be calculated using the scattering the scalar field's coefficient. In addition, a time-dependent for TLNs measured by a comoving observer is obtained. 

In this paper, we will calculate tidal response of charged and uncharged AdS black holes using scalar field perturbation. To obtain analytical results, we perform near-horizon approximation to radial Klein-Gordon equation. In section \ref{sec3}, we investigate tidal response of uncharged black holes including static Schwarzschild AdS and rotating Kerr AdS. Then, we consider charged black holes consisting of static charged Reissner-Nordstrom AdS and rotating charged Kerr-Newman AdS in section \ref{sec4}.

\section{Tidal response of an object due to external tidal field}\label{sec2}
When a spherically symmetric object is placed in a free-field empty space, it has only monopole moment. Then, if an external field presents, the object is deformed shown by the emergence or change of its multipole moments.The deformation occurs inside the object is called tidal response of a self-gravitating object caused by the external tidal field. According to Newtonian formalism, potential at distance $r$ from the center of mass of the object of mass $M$ is written as \citep{poisson2014gravity,le2021tidal}
\begin{equation}\label{eq:01}
	U=\frac{M}{r}+\sum\limits_{l=2}^\infty \sum\limits_{m=-l}^l\left[\frac{(2l-1)!!}{l!}\frac{I_{lm}Y_{lm}}{r^{l+1}}-\frac{(l-2)!}{l!}\epsilon_{lm}Y_{lm}r^{l}\right].
\end{equation} 
The first and second term stand for the potential created by the object and the external tidal field, respectively. $I_{lm}$ is the induced multipole moments on the object caused by the external tidal field $\epsilon_{lm}$.

In the linear approximation, $I_{lm}$ and $\epsilon_{lm}$ are related by the following equation \citep{bhatt2024response}
\begin{equation}\label{eq:02}
	I_{lm}=\frac{(l-2)!}{(2l-1)!!}R^{2l+1}\left[2k_{lm}\epsilon_{lm}-\tau_{0}\nu_{lm}\dot{\epsilon}_{lm}+...\right].
\end{equation}
$R$ is the radial scale of the object and $k_{lm}$ is the TLNs. Moreover, $\nu_{lm}$ is the TDs while $\tau_{0}$ is the viscocity induced time delay denoting the time between the external tidal field and the object's response.
Combining equations \ref{eq:01} and \ref{eq:02} leads to the following expression
\begin{equation}\label{eq:03}
	U=\frac{M}{r}+\sum\limits_{l=2}^\infty \sum\limits_{m=-l}^l\frac{(2l-1)!}{l!}\epsilon_{lm}r^{lm}\left[1+F_{lm}(\omega)\left(\frac{R}{r}\right)^{2l+1}\right]Y_{lm}.
\end{equation}
$F_{lm}$ is defined as the tidal response
\begin{equation}\label{eq:04}
	F_{lm}=2k_{lm}+i\omega \tau_{0}\nu_{lm}+O(\omega^2).
\end{equation}
From equation \ref{eq:03}, we can identify tidal response as the quantity which is proportional to the ratio between the decaying and growing part of the potential.

In the Newman-Penrose formalism, there is a relation between scalar and gravitational potential in the Newtonian domain. From that relation, tidal response can be determined by considering the behaviour of the scalar field in the intermediate region \citep{le2021tidal}. Adopting the same method, we can compute tidal response of the compact object perturbed by the scalar field as follow \citep{teukolsky1972rotating,creci2021tidal}
\begin{equation}\label{eq:05}
	k_{lm}=\frac{\mathrm{coefficient\hspace{1mm}of\hspace{1mm}the\hspace{1mm}decaying\hspace{1mm}part\hspace{1mm}of\hspace{1mm}the\hspace{1mm}field\hspace{1mm}\phi}}{\mathrm{coefficient\hspace{1mm}of\hspace{1mm}the\hspace{1mm}growing\hspace{1mm}part\hspace{1mm}of\hspace{1mm}the\hspace{1mm}field}\hspace{1mm}\phi}.
\end{equation}

\section{Tidal response of uncharged black holes}\label{sec3}
The tidal response of uncharged black holes can be calculated using massless and neutral scalar field perturbation. In general relativity, the dynamic of scalar field $\Phi$ is governed by Klein-Gordon equation on curved spacetime which can be written as
\begin{equation}\label{eq:1}
	\frac{1}{\sqrt{-g}}\partial_{\mu}\left[\sqrt{-g}g^{\mu\nu}\partial_\nu\Phi\right]=0.
\end{equation}
$g^{\mu\nu}$ denotes the contravariant metric tensor and $g$ stands for the determinant of spacetime metric. Klein-Gordon equation is known to be separable, so to obtain its solution we can take the following ansatz 
\begin{equation}\label{eq:2}
	\Phi(t,r,\theta,\phi)=e^{-i\omega t}e^{im\phi}S(\theta)R(r).
\end{equation}
In this section, we will consider Schwarzschild AdS and Kerr AdS black holes.
\subsection{Schwarzschild AdS}\label{2.1}
We begin our calculation from the simplest solution of Einstein field equation in AdS space known as Schwarzschild AdS black hole. The metric of such black hole is given by
\begin{equation}\label{eq:3}
	ds^2=-\left(1-\frac{2M}{r}+\frac{r^2}{l_c^2}\right)dt^2+\frac{dr^2}{1-\frac{2M}{r}+\frac{r^2}{l_c^2}}+r^2 d\theta^2+r^2\sin^2{\theta} d\phi^2,
\end{equation}
where $M$ and $l_c$ are the black hole's mass and AdS radius, respectively.
The angular and radial components of Klein-Gordon equation satisfy
\begin{equation}\label{eq:4}
	\begin{split}
		\frac{1}{\sin\theta S(\theta)}\frac{d}{d\theta}\left[\sin\theta\frac{d}{d\theta}S(\theta)\right]-\frac{m^2}{\sin^2{\theta}}&=-l(l+1)\\
		\frac{1}{R(r)}\frac{d}{dr}\left[\Delta\frac{d}{dr}R(r)\right]-\left[l\left(l+1\right)-\frac{\omega^2r^4}{\Delta}\right]&=0.
	\end{split}
\end{equation}
Here, we define $\Delta=r^2f$ which is a quantity determining the location of event horizon
\begin{equation}\label{eq:5}
	\Delta=\frac{r^4}{l_c^2}+2Mr+r^2.
\end{equation}

The radial part of equation \ref{eq:4} is difficult to solve analytically. Therefore, we use near-horizon approximation $r/\Delta\approx r_{+}/\Delta$ as used in Ref \citep{charalambous2022love} and perform Taylor expansion for $\Delta$ near $r=r_{+}$
\begin{equation}\label{eq:6}
	\begin{split}
		\Delta&\approx\Delta(r_{+})+\left(r-r_{+}\right)\Delta'(r_{+})+\frac{\left(r-r_{+}\right)}{2!}\Delta(r_{+})''\\
		\Delta&\approx k_{Sch}\left(r-r_{+}\right)\left(r-r_{*}\right),
	\end{split}	
\end{equation}
where $k_{Sch}$ and $r_{*}$ are written as
\begin{equation}\label{eq:7}
	k_{Sch}=1+\frac{6r_{+}^2}{l_{c}^2} ; r_{*}=r_{+}-\frac{2r_{+}-2M+4\frac{r_{+}^3}{l_{c}^2}}{1+\frac{6r_{+}^2}{l_{c}^2}}.
\end{equation}
Consequently, equation \ref{eq:4} transforms to
\begin{equation}\label{eq:8}
	\frac{d}{dr}\left[(r-r_{+})(r-r_{*})\frac{dR(r)}{dr}\right]-\left[\frac{l(l+1)}{k_{Sch}}-\frac{\omega^2r_{+}^4}{k_{Sch}^2(r-r_{+})(r-r_{*})}\right]R(r)=0.
\end{equation}
To solve equation \ref{eq:8}, we make the following coordinate transformation
\begin{equation}\label{eq:9}
	z=\frac{r-r_{+}}{r-r_{*}},
\end{equation}
and we have
\begin{equation}\label{eq:10}
	\begin{split}
		&z(1-z)\frac{d^{2}R(r)}{dz^2}+(1-z)\frac{dR(r)}{dz}\\&+\left[\frac{\omega^2r_{+}^4}{k_{Sch}^2(r_{+}-r_{-})^2z}-\frac{\omega^2r_{+}^4}{k_{Sch}^2(r_{+}-r_{-})^2}-\frac{l(l+1)}{k_{Sch}(1-z)}\right]R(r)=0.
	\end{split}
\end{equation}

To calculate tidal response, we need to consider the ingoing solution of equation \ref{eq:10} and using Appendix \ref{A1} we obtain 
\begin{equation}\label{eq:11}
	R_{in}(z)=k_{1}z^{-i\sqrt{A}}\left(1-z\right)^{L+1}{}_2 F_1\left(a_s,b_s;c_s;z\right)
\end{equation}
$L$,$a_s$,$b_s$, and $c_s$ are
\begin{equation}\label{eq:12}
	\begin{split}
		L&=\frac{-1+\sqrt{1+\frac{4l(l+1)}{k_{Sch}}}}{2},\hspace{4mm} a_s=1+L-2i\frac{\omega r_{+}^2}{k_{Sch}(r_{+}-r_{-})}\\
		b_s&=1+L,\hspace{4mm} c_s=1-2i\frac{\omega r_{+}^2}{k_{Sch}(r_{+}-r_{-})}.\\	
	\end{split}
\end{equation}
Using hypergeometric relation given in Appendix \ref{A2}, the ingoing solution can be decomposed as follow
\begin{equation}\label{eq:13}
	\begin{split}	
		&R_{in}(z)=k_{1}z^{-i\sqrt{A}}\left[\left(1-z\right)^{L+1}\frac{\Gamma(c_s)\Gamma(c_s-a_s-b_s)}{\Gamma(c_s-a_s)\Gamma(c_s-a_s)}{}_2F_1\left(a_s,b_s;a_s+b_s-c_s+1;1-z\right)\right. \\&\left.+\left(1-z\right)^{-L}
		\frac{\Gamma(c_s)\Gamma(a_s+b_s-c_s)}{\Gamma(a_s)\Gamma(b_s)}
		{}_2F_1\left(c_s-a_s,c_s-b_s;c_s-b_s+1;1-z\right)\right].
	\end{split}
\end{equation}
Hence, using the definition in equation \ref{eq:05} we get the tidal response
\begin{equation}\label{eq:15}
	\begin{split}
		k_{lm}&=\frac{\Gamma(c_s-a_s-b_s)\Gamma(a_s)\Gamma(b_s)}{\Gamma(c_s-a_s)\Gamma{c_s-b_s}\Gamma(a_s+b_s-c_s)}\\
		k_{lm}&=\frac{\Gamma(-2L-1)\Gamma(1+L)\Gamma\left(1+L-2i\frac{\omega r_{+}^2}{k_{Sch}(r_{+}-r_{-})}\right)}{\Gamma(-L)\Gamma(2L+1)\Gamma\left(-L-2i\frac{\omega r_{+}^2}{k_{Sch}(r_{+}-r_{-})}\right)}.
	\end{split}
\end{equation}	
From the relation of gamma function derived in \ref{A3} we can identify the TLNs $\left(\kappa_{lm}\right)$ and TDs $\left(\nu_{lm}\right)$ of Schwarzschild black hole
\begin{equation}\label{eq:16}
	\begin{split}
		\kappa_{lm}&=-\frac{\Gamma(-2L-1)\Gamma(L+1)}{\Gamma(-L)\Gamma(2L+1)}\sin\left(\pi L\right)\cosh\left[\frac{2\pi\omega r_{+}^2}{k_{Sch}(r_{+}-r_{-})}\right]\frac{\left|\Gamma\left(1+L-2i\frac{\omega r_{+}^2}{k_{Sch}(r_{+}-r_{-})}\right)\right|^2}{\pi}\\
		\nu_{lm}&=-\frac{\Gamma(-2L-1)\Gamma(L+1)}{\Gamma(-L)\Gamma(2L+1)}\cos\left(\pi L\right)\sinh\left[\frac{2\pi\omega r_{+}^2}{k_{Sch}(r_{+}-r_{-})}\right]\frac{\left|\Gamma\left(1+L-2i\frac{\omega r_{+}^2}{k_{Sch}(r_{+}-r_{-})}\right)\right|^2}{\pi}.
	\end{split}
\end{equation}

Variable $L$ is generally non-integer due to the existence of cosmological constant represented by AdS radius $l_{c}^2$ and it makes TLNs never vanish. On the other hand, tidal dissipation can vanish for static scalar field perturbation $\omega=0$. For this case, TLNs and tidal dissipation are
\begin{equation}\label{eq:17}
	\begin{split}
		\kappa_{lm}&=-\frac{\Gamma(-2L-1)\Gamma(L+1)}{\Gamma(-L)\Gamma(2L+1)}\sin\left(\pi L\right)\frac{\left|\Gamma\left(1+L\right)\right|^2}{\pi}\\
		\nu_{lm}&=0.
	\end{split}
\end{equation}.

\subsection{Kerr AdS}
In this subsection, we will study the tidal response of rotating black hole in AdS space namely Kerr AdS black hole. This solution is derived from Einstein field equation with a negative cosmological constant. The metric can be expressed as 
\begin{equation}\label{eq:18}
	\begin{split}
		ds^2=&-\frac{\Delta_{r}}{\Sigma}\left(dt-\frac{a\sin^2{\theta}}{\Xi}d\phi\right)^{2}+\frac{\Sigma}{\Delta_{r}} dr^{2}+\frac{\Sigma}{\Delta_{\theta}}d\theta^2\\
		&+\frac{\Delta_{\theta}}{\Sigma}\sin^2{\theta}\left(a dt-\frac{(r^2+a^2)}{\Xi}d\phi\right)^2,
	\end{split}
\end{equation}
where $\Delta_{r}$, $\Delta_{\theta}$, $\Sigma$, and $\Xi$ are
\begin{equation}\label{eq:19}
	\begin{split}
		\Delta_{r}=&\left(r^2+a^2\right)\left(1+\frac{r^2}{l_c^2}\right)-2Mr,\hspace{3mm}\Delta_{\theta}=1-\frac{a^2}{l_c^2}\cos^2{\theta}\\
		\Sigma=&r^2+a^2\cos^2{\theta},\hspace{3mm}\Xi=1-\frac{a^2}{l_c^2}.
	\end{split}
\end{equation}
$a$ stands for the angular momentum per unit mass of the black hole. Taylor expansion for $\Delta_{r}$ near $r=r_{+}$ is
\begin{equation}\label{eqq:19}
	\Delta\approx k_{\mathrm{Kerr}}\left(r-r_{+}\right)\left(r-r_{*}\right),
\end{equation}
where
\begin{equation}\label{eqqq:19}
	\begin{split}
		k_{\mathrm{Kerr}}&=1+\frac{a^2}{l_{c}^2}+6\frac{r_{+}^2}{l_{c}^2},\\
		r_{*}&=r_{+}-\frac{1}{k_{\mathrm{Kerr}}r_{+}}\left[r_{+}^2-a^2+\frac{3r_{+}^4}{l_{c}^2}+\frac{a^2r_{+}^2}{l_{c}^2}\right].
	\end{split}
\end{equation}

Plugging contravariant metric tensor of equation \ref{eq:18} to \ref{eq:1} yields the angular and radial component of Klein-Gordon equation
\begin{equation}\label{eq:20}
	\begin{split}
		\frac{1}{\sin\theta}\frac{d}{d\theta}\left[\Delta_{\theta}\sin\theta\frac{dS(\theta)}{d\theta}\right]-\left[\frac{\left(\omega a\sin\theta-\frac{m\Xi}{\sin\theta}\right)^2}{\Delta_{\theta}}+\Lambda_{lm}\right]S(\theta)&=0\\
		\frac{d}{dr}\left[\Delta \frac{dR(r)}{dr}\right]+\left(\frac{\left(\omega(r^2+a^2)-ma\Xi\right)^2}{\Delta_{r}}-\Lambda_{lm}\right)R(r)&=0,
	\end{split}
\end{equation}
where $\Lambda_{lm}$ is the separation constant.
Employing the same near-horizon approximation and coordinate transformation as in the previous subsection gives the transformed radial component of Kerr AdS
\begin{equation}\label{eq:21}
	\begin{split}
		&\frac{d}{dr}\left[(r-r_{+})(r-r_{*})\frac{d}{dr}R\right]-\left[\frac{\Lambda_{lm}}{k_{\mathrm{Kerr}}}-\frac{\left[\omega \left(r_{+}^2+a^2\right)-ma\Xi\right]^2}{k_{\mathrm{Kerr}}^2(r-r_{+})(r-r_{*})}\right]R(r)=0\\
		&z(1-z)\frac{d^{2}R(r)}{dz^2}+(1-z)\frac{dR(r)}{dz}\\&+\left[\frac{\left(\omega(r^2+a^2)-ma\Xi\right)^2}{k_{\mathrm{Kerr}}^2(r_{+}-r_{-})^2z}-\frac{\left(\omega(r^2+a^2)-ma\Xi\right)^2}{k_{\mathrm{Kerr}}^2(r_{+}-r_{-})^2}-\frac{\Lambda_{lm}}{k_{\mathrm{Kerr}}(1-z)}\right]R(r)=0.	
	\end{split}
\end{equation}
Following the same steps as in section \ref{2.1}, we obtain TLNs $\left(\kappa_{lm}\right)$ and TDs $\left(\nu_{lm}\right)$ of Kerr AdS

\begin{equation}\label{eq:22}
	\begin{split}
		\kappa_{lm}=&-\frac{\Gamma(-2L-1)\Gamma(L+1)}{\Gamma(-L)\Gamma(2L+1)}\sin\left(\pi L\right)\cosh\left[2\pi\frac{\omega \left(r_{+}^2+a^2\right)-ma\Xi}{k_{\mathrm{Kerr}}(r_{+}-r_{*})}\right]\\
		&\times\frac{\left|\Gamma\left(1+L-2i\frac{\omega \left(r_{+}^2+a^2\right)-ma\Xi}{k_{\mathrm{Kerr}}(r_{+}-r_{*})}\right)\right|^2}{\pi}\\
		\nu_{lm}=&-\frac{\Gamma(-2L-1)\Gamma(L+1)}{\Gamma(-L)\Gamma(2L+1)}\cos\left(\pi L\right)\sinh\left[2\pi\frac{\omega \left(r_{+}^2+a^2\right)-ma\Xi}{k_{\mathrm{Kerr}}(r_{+}-r_{*})}\right]\\
		&\times\frac{\left|\Gamma\left(1+L-2i\frac{\omega \left(r_{+}^2+a^2\right)-ma\Xi}{k_{\mathrm{Kerr}}(r_{+}-r_{*})}\right)\right|^2}{\pi}.
	\end{split}
\end{equation}
In the case of Kerr AdS, $L$ is given by
\begin{equation}
	L=\frac{-1+\sqrt{1+\frac{4\Lambda_{lm}}{k_{\mathrm{Kerr}}}}}{2}.	
\end{equation}

Let us consider special cases for the condition of the scalar field and its influence to the tidal dissipation:
\begin{itemize}
	\item In the case of static scalar field $\omega=0$, $\nu_{lm}$ has a nonzero value due to the angular momentum per unit mass $a$. Physically, it can be explained that spinning black hole causes frame dragging effect yielding dissipation to the scalar field.
	\item For static $\omega=0$ and axisymmetric scalar field $m=0$,
	$\nu_{lm}=0$. This case is similar to static perturbation on Schwarzschild AdS. 
	\item For special case $\omega \left(r_{+}^2+a^2\right)=ma\Xi$, we find $\nu_{lm}=0$. It implies that there is a specific case where frame dragging effect is balanced by oscillating scalar field to avoid dissipation.
\end{itemize}

\section{Tidal response of charged black holes}\label{sec4}
In this section, we will calculate tidal response of black holes carrying electric charge $Q$. Consquently, we can perturb black holes using charged scalar field. The covariant derivative transformas as
$\partial_{\mu}\rightarrow D_{\mu}=\partial_{\mu}-iqA_{\mu}$, and therefore Klein-Gordon equation becomes
\begin{equation}
	\begin{split}
		D_{\mu}\left(\sqrt{-g}g^{\mu\nu}D_{\nu}\Phi\right)&=0\\
		\partial_{\mu}\left(\sqrt{-g}g^{\mu\nu}\partial_{\nu}\Phi\right)
		-iq\partial_{\mu}\left(\sqrt{-g}g^{\mu\nu}A_{\nu}\Phi\right)
		\\-iq A_{\mu}\left(\sqrt{-g}g^{\mu\nu}\partial_{\nu}\Phi\right)
		-q^2A_{\mu}\sqrt{-g}g^{\mu\nu}A_{\nu}\Phi&=0.
	\end{split}	
\end{equation}
Here, $q$ and $A_{\mu}$ are the charge of scalar field and electromagnetic potential of the black hole, respectively. In this section, we will consider Reissner-Nordstrom AdS and Kerr-Newman AdS black holes.

\subsection{Reissner-Nordstrom AdS}
The spacetime metric and electromagnetic potential of Reissner-Nordstrom AdS take the following form 
\begin{equation}
	\begin{split}
		ds^2&=-\left(1-\frac{2M}{r}+\frac{Q^2}{r^2}+\frac{r^2}{l_c^2}\right)dt^2+\frac{dr^2}{1-\frac{2M}{r}+\frac{Q^2}{r^2}+\frac{r^2}{l_c^2}}+r^2 d\theta^2\\&+r^2\sin^2{\theta} d\phi^2, \hspace{3mm} A_{\mu}dx^{\mu}=-\frac{Q}{r}dt.
	\end{split}
\end{equation}
The angular and radial components of Klein-Gordon equation are
\begin{equation}
	\begin{split}
		\frac{1}{\sin\theta S(\theta)}\frac{d}{d\theta}\left[\sin\theta\frac{d}{d\theta}S(\theta)\right]-\frac{m^2}{\sin^2{\theta}}&=-l(l+1)\\
		\frac{d}{dr}\left[\Delta\frac{dR(r)}{dr}\right]-\left[l(l+1)-\frac{(\omega r^2-qQr)^2}{\Delta}\right]R(r)&=0.
	\end{split}
\end{equation}
Taylor expansion for $\Delta$ near $r=r_{+}$ can be expressed as
\begin{equation}
	\Delta\approx k_{\mathrm{RN}}\left(r-r_{+}\right)\left(r-r_{*}\right),
\end{equation}
where
\begin{equation}
	\begin{split}
		k_{\mathrm{RN}}&=1+6\frac{r_{+}^2}{l_{c}^2},\\
		r_{*}&=r_{+}-\frac{1}{k_{\mathrm{RN}}}\left[2r_{+}-2M+\frac{4r_{+}^3}{l_{c}^2}\right].
	\end{split}
\end{equation}
The transformed radial equation of Reissner-Nordstrom AdS is written as
\begin{equation}
	\begin{split}
		&\frac{d}{dr}\left[(r-r_{+})(r-r_{*})\frac{d}{dr}R\right]-\left[\frac{l(l+1)}{k_{\mathrm{RN}}}-\frac{(\omega r_{+}^2-qQr_{+})^2}{k_{\mathrm{RN}}^2(r-r_{+})(r-r_{*})}\right]R(r)=0\\
		&z(1-z)\frac{d^{2}R(r)}{dz^2}+(1-z)\frac{dR(r)}{dz}\\&+\left[\frac{\left(\omega r_{+}^2-qQr_{+}\right)^2}{k_{\mathrm{RN}}^2(r_{+}-r_{-})^2z}-\frac{\left(\omega r_{+}^2-qQr_{+}\right)^2}{k_{\mathrm{RN}}^2(r_{+}-r_{-})^2}-\frac{l(l+1)}{k_{\mathrm{RN}}(1-z)}\right]R(r)=0.
	\end{split} 
\end{equation} 
TLNs $\left(\kappa_{lm}\right)$ and TDs $\left(\nu_{lm}\right)$ of Reissner-Nordstrom AdS are given by
\begin{equation}
	\begin{split}
		\kappa_{lm}=&-\frac{\Gamma(-2L-1)\Gamma(L+1)}{\Gamma(-L)\Gamma(2L+1)}\sin\left(\pi L\right)\cosh\left[2\pi\frac{\omega r_{+}^2-qQr_{+}}{k_{\mathrm{RN}}(r_{+}-r_{-})}\right]\\
		&\times\frac{\left|\Gamma\left(1+L-2i\frac{\omega r_{+}^2-qQr_{+}}{k_{\mathrm{RN}}(r_{+}-r_{-})}\right)\right|^2}{\pi}\\
		\nu_{lm}=&-\frac{\Gamma(-2L-1)\Gamma(L+1)}{\Gamma(-L)\Gamma(2L+1)}\cos\left(\pi L\right)\sinh\left[2\pi\frac{\omega r_{+}^2-qQr_{+}}{k_{\mathrm{RN}}(r_{+}-r_{-})}\right]\\
		&\times\frac{\left|\Gamma\left(1+L-2i\frac{\omega r_{+}^2-qQr_{+}}{k_{\mathrm{RN}}(r_{+}-r_{-})}\right)\right|^2}{\pi}
	\end{split}
\end{equation}
In the case of Reissner-Nordstrom AdS, $L$ is written as
\begin{equation}
	L=\frac{-1+\sqrt{1+\frac{4l(l+1)}{k_{\mathrm{RN}}}}}{2}.	
\end{equation}
Specific condition of the scalar field and its influence to the tidal dissipation
\begin{itemize}
	\item In the case of static scalar field $\omega=0$, $\nu_{lm}$ has a nonzero value due to the Coulomb interaction between the field and the black hole. 
	\item For special case $\omega r_{+}^2=qQ$, we find $\nu_{lm}=0$. It implies that there is a specific case where Coulomb interraction is balanced by oscillating scalar field to make dissipation vanishes.
\end{itemize}

\subsection{Kerr-Newman AdS}
In this subsection, we will consider the most general solution of Einstein field equation in AdS space namely Kerr-Newman AdS black hole. Not only does black hole rotate but also contains electric charge. The spacetime metric of such black hole can be expressed as
\begin{equation}
	\begin{split}
		ds^2=&-\frac{\Delta_{r}}{\Sigma}\left(dt-\frac{a\sin^2{\theta}}{\Xi}d\phi\right)^2+\frac{\Sigma}{\Delta_r}dr^2+\frac{\Sigma}{\Delta_{\theta}}d\theta^2\\
		&+\frac{\Delta_{\theta}\sin^2{\theta}}{\Sigma}\left[a dt-\frac{r^2+a^2}{\Xi}d\phi\right]^2,
	\end{split}
\end{equation}
where $\Delta_{\theta}$, $\Sigma$, and $\Xi$ are the same as that of Kerr AdS, while $\Delta_r$ is
\begin{equation}
	\Delta_{r}=\left(r^2+a^2\right)\left(1+\frac{r^2}{l_c^2}\right)-2Mr+Q^2.
\end{equation}
The angular and radial components of Klein-Gordon equation are
\begin{equation}
	\begin{split}
		\frac{1}{\sin\theta}\frac{d}{d\theta}\left[\Delta_{\theta}\sin\theta\frac{dS(\theta)}{d\theta}\right]-\left[\frac{\left(\omega a\sin\theta-\frac{m\Xi}{\sin\theta}\right)^2}{\Delta_{\theta}}+\Lambda_{lm}\right]S(\theta)&=0\\
		\frac{d}{dr}\left[\Delta \frac{dR(r)}{dr}\right]+\left(\frac{\left(\omega(r^2+a^2)-ma\Xi-qQr\right)^2}{\Delta_{r}}-\Lambda_{lm}\right)R(r)&=0.
	\end{split}
\end{equation}
Taylor expansion for $\Delta_{r}$ near $r=r_{+}$ is given by
\begin{equation}
	\Delta\approx k_{\mathrm{KN}}\left(r-r_{+}\right)\left(r-r_{*}\right),
\end{equation}
where
\begin{equation}
	\begin{split}
		k_{\mathrm{KN}}&=1+\frac{a^2}{l_{c}^2}+6\frac{r_{+}^2}{l_{c}^2},\\
		r_{*}&=r_{+}-\frac{1}{k_{\mathrm{KN}}r_{+}}\left[r_{+}^2-a^2+\frac{3r_{+}^4}{l_{c}^2}+\frac{a^2r_{+}^2}{l_{c}^2}\right].
	\end{split}
\end{equation}
The transformed radial equation of Kerr-Newman AdS is written as
\begin{equation}
	\begin{split}
		&\frac{d}{dr}\left[(r-r_{+})(r-r_{*})\frac{d}{dr}R\right]-\left[\frac{\Lambda_{lm}}{k_{\mathrm{KN}}}-\frac{\left[\omega \left(r_{+}^2+a^2\right)-ma\Xi-qQr_{+}\right]^2}{k_{\mathrm{KN}}^2(r-r_{+})(r-r_{*})}\right]R(r)=0\\
		&z(1-z)\frac{d^{2}R(r)}{dz^2}+(1-z)\frac{dR(r)}{dz}+\left[\frac{\left(\omega(r^2+a^2)-ma\Xi-qQr\right)^2}{k_{\mathrm{KN}}^2(r_{+}-r_{-})^2z}\right. \\&\left.-\frac{\left(\omega(r^2+a^2)-ma\Xi-qQr_{+}\right)^2}{k_{\mathrm{KN}}^2(r_{+}-r_{-})^2}-\frac{l(l+1)}{k_{\mathrm{KN}}(1-z)}\right]R(r)=0.
	\end{split}
\end{equation}
In the case of Kerr-Newman AdS, $L$ is given by
\begin{equation}
	L=\frac{-1+\sqrt{1+\frac{4\Lambda_{lm}}{k_{\mathrm{KN}}}}}{2}.	
\end{equation}
TLNs $\left(\kappa_{lm}\right)$ and TDs $\left(\nu_{lm}\right)$ of Kerr-Newman AdS are 
\begin{equation}
	\begin{split}
		\kappa_{lm}=&-\frac{\Gamma(-2L-1)\Gamma(L+1)}{\Gamma(-L)\Gamma(2L+1)}\sin\left(\pi L\right)\cosh\left[2\pi\frac{\omega \left(r_{+}^2+a^2\right)-ma\Xi-qQr_{+}}{k_{\mathrm{KN}}(r_{+}-r_{*})}\right]\\
		&\times\frac{\left|\Gamma\left(1+L-2i\frac{\omega \left(r_{+}^2+a^2\right)-ma\Xi-qQr_{+}}{k_{\mathrm{KN}}(r_{+}-r_{*})}\right)\right|^2}{\pi}\\
		\nu_{lm}=&-\frac{\Gamma(-2L-1)\Gamma(L+1)}{\Gamma(-L)\Gamma(2L+1)}\cos\left(\pi L\right)\sinh\left[2\pi\frac{\omega \left(r_{+}^2+a^2\right)-ma\Xi-qQr_{+}}{k_{\mathrm{KN}}(r_{+}-r_{*})}\right]\\
		&\times\frac{\left|\Gamma\left(1+L-2i\frac{\omega \left(r_{+}^2+a^2\right)-ma\Xi-qQr_{+}}{k_{\mathrm{KN}}(r_{+}-r_{*})}\right)\right|^2}{\pi}.
	\end{split}
\end{equation}
Specific condition of the scalar field and its influence to the tidal dissipation
\begin{itemize}
	\item In the case of static scalar field $\omega=0$, $\nu_{lm}$ has a nonzero value due to the Coulomb interraction and rotating motion of the black hole. 
	\item For static $\omega=0$ and axisymmetric scalar field $m=0$,
	$\nu_{lm}$ has a nonzero value because Coulomb interraction prevents it from vanishing.
	\item For special case $\omega \left(r_{+}^2+a^2\right)=ma\Xi+qQr_{+}$, we find $\nu_{lm}=0$. It implies that there is a specific case where Coulomb interraction and frame dragging are balanced by oscillating scalar field to make dissipation vanishes.
\end{itemize}

\begin{longtable}{|l|ll|ll|ll|}
	\hline
	\multicolumn{1}{|c|}{\multirow{2}{*}{Black holes}} &
	\multicolumn{2}{l|}{\begin{tabular}[c]{@{}c@{}}Scalar field with \\ frequency $\omega$\end{tabular}} &
	\multicolumn{2}{l|}{\begin{tabular}[c]{@{}c@{}}Static scalar field\\ $\left(\omega=0\right)$\end{tabular}} &
	\multicolumn{2}{l|}{\begin{tabular}[c]{@{}c@{}}Special case \\ $\omega r_{+}=qQ$\end{tabular}} \\   
	\multicolumn{1}{|c|}{} &
	\multicolumn{1}{c|}{TLNs} &
	\multicolumn{1}{c|}{TDs} &
	\multicolumn{1}{c|}{TLNs} &
	\multicolumn{1}{c|}{TDs} &
	\multicolumn{1}{c|}{TLNs} &
	\multicolumn{1}{c|}{TDs} \\ \hline
	\endfirsthead
	\endhead
	\begin{tabular}[c]{@{}l@{}}Schwarzschild \\ AdS\end{tabular}      & \multicolumn{1}{c|}{\checkmark} & 
	\multicolumn{1}{c|}{\checkmark}  
	&  
	\multicolumn{1}{c|}{\checkmark} & 
	\multicolumn{1}{c|}{0}  & 
	\multicolumn{1}{c|}{-} & 
	\multicolumn{1}{c|}{-} 
	\\ \hline
	\begin{tabular}[c]{@{}l@{}}Reissner-Nordstrom \\ AdS\end{tabular} & \multicolumn{1}{c|}{\checkmark} & 
	\multicolumn{1}{c|}{\checkmark}  
	&  
	\multicolumn{1}{c|}{\checkmark} & 
	\multicolumn{1}{c|}{\checkmark} 
	& 
	\multicolumn{1}{c|}{\checkmark} & 
	\multicolumn{1}{c|}{0} 
	\\ \hline
	\caption{TLNs and TDs of static AdS black holes
	}
	\label{table1}\\
\end{longtable}
\begin{longtable}{|l|cc|cc|cc|cl|}
	\hline
	\multicolumn{1}{|c|}{\multirow{2}{*}{Black holes}} &
	\multicolumn{2}{c|}{\begin{tabular}[c]{@{}c@{}}Scalar field with \\ frequency $\omega$\end{tabular}} &
	\multicolumn{2}{c|}{\begin{tabular}[c]{@{}c@{}}Static scalar \\ field $\left(\omega=0\right)$\end{tabular}} &
	\multicolumn{2}{c|}{\begin{tabular}[c]{@{}c@{}}Static and \\ axisymmetric \\
			$\left(\omega=m=0\right)$
	\end{tabular}} &
	\multicolumn{2}{l|}{\begin{tabular}[c]{@{}c@{}}Special case
	\end{tabular}} \\ 
	\multicolumn{1}{|c|}{} &
	\multicolumn{1}{c|}{TLNs} &
	TDs &
	\multicolumn{1}{c|}{TLNs} &
	TDs &
	\multicolumn{1}{c|}{TLNs} &
	TDs &
	\multicolumn{1}{c|}{TLNs} &
	TDs \\ \hline
	\endfirsthead
	\endhead
	Kerr AdS &
	\multicolumn{1}{c|}{
		\checkmark} &
	\checkmark &
	\multicolumn{1}{c|}{\checkmark} &
	\checkmark &
	\multicolumn{1}{c|}{\checkmark} &
	0 &
	\multicolumn{1}{c|}{\checkmark} &
	\multicolumn{1}{c|}{0} \\ \hline
	\begin{tabular}[c]{@{}l@{}}Kerr-Newman \\ AdS\end{tabular} &
	\multicolumn{1}{c|}{\checkmark} &
	\checkmark &
	\multicolumn{1}{c|}{\checkmark} &
	\checkmark&
	\multicolumn{1}{c|}{\checkmark} &
	\checkmark &
	\multicolumn{1}{c|}{\checkmark} &
	\multicolumn{1}{c|}{0} \\ \hline
	\caption{TLNs and TDs of rotating AdS black holes
	}
	\label{table2}\\
\end{longtable}

Our results for TLNs and TDs of static and rotating black holes are summarized in the above tables \ref{table1} and \ref{table2}. The sign $"\checkmark"$ denotes nonzero value of a quantity while $"0"$ stands for vanishing one.

\section{Discussion}
The tidal response of static and rotating black holes in AdS spacetime has been investigated using scalar field perturbation. For uncharged and charged black holes, we perturb them using neutral and charged scalar field, respectively. Our results show that TLNs never vanish due to the presence of cosmological constant represented by AdS radius $l_c$. Regardless of the condition of the scalar field and black hole's parameters, TLNs always has nonzero value. However, tidal dissipation can vanish in various conditions for different black holes.

Comparing our results to that that of general relativity, we show that black holes can be deformed by scalar field perturbation if cosmological constant is taken into account. We argue that nonzero TLNs provide a better description to the black hole's geometry rather than think of it as the most robust object in the universe. Because, accomodating cosmological constant is essential to explain why our universe is expanding. Despite its role as the dark energy, it has a significant impact to the dynamics of the astrophysical objects. Therefore, the influence of cosmological constant to the dynamics of other stellar objects such as neutron stars and white dwarfs needs further investigation.

Since Schwarzschild and Kerr black holes are uncharged, we consider a neutral scalar field. It has been found that tidal dissipation of Schwarzschild black hole is zero for static tidal perturbation. Meanwhile, Kerr black hole requires additional condition to have vanishing tidal dissipation. Not only is the scalar field static, but it has also to be axisymmetric. Moreover, there is a specific relation between scalar field's frequency and angular momentum per unit mass which leads to zero tidal dissipation.

In the case of charged black holes i.e. Reissner-Nordstrom and Kerr-Newman, the charged scalar field is considered to study the tidal response of such black holes. Tidal dissipation is nonzero for static scalar field showing that Coulomb interaction between scalar field and black hole causes a dissipative effect to the spacetime background around the black hole. Nevertheless, Reissner-Nordstrom black hole can have vanishing tidal dissipation when the scalar field and electric charge satisfy specific relation. Interestingly, the existence of tidal dissipation of Kerr-Newman black hole is neither influenced by zero value of scalar field's frequency or its axisymmetric property. But, it can be vanish when the scalar field's frequency, angular momentum per unit mass, and electric charge fulfill certain relation. 

TLNs of Reissner-Nordstrom and Schwarzschild AdS are similar if those black holes are perturbed using neutral scalar field. They are distinguished in magnitude only by the electric charge of Reissner-Nordstrom black hole. Perturbing by the same field, the same result holds for Kerr-Newman and Kerr AdS as well. TLNs are similar although Coulomb interaction does not present. Moreover, TLNs for rotating and static black holes are distingusihed only by rotation parameter.

Another interesting result stemming from our TLNs calculation of the charged black holes is the absence of discontinuity between neutral and charged scalar field as appears in \citep{ma2025charging}. Even if the neutral scalar is used to perturb the black holes, TLNs do not vanish and are similar to that of uncharged counterparts. It implies that the discontinuity issue of TLNs of black hole derived from Einstein-Maxwell theory can be resolved by considering cosmological constant. The calculation of tidal response of AdS black holes using other methods such as effective field theory and Ernst formalism deserves to be explored.

\backmatter

\bmhead{Acknowledgements}
F. P. Z. would like to thank Fakultas
Matematika dan Ilmu Pengetahuan Alam, Institut Teknologi Bandung and the Ministry of
Higher Education, Science, and Technology (Kemendikti Saintek) for partial financial support. Yusmantoro would like to thank Indonesia Endowment Fund for Education Agency (Lembaga Pengelola Dana Pendidikan/LPDP).

\section*{Declaration}
The authors declare no competing interests.


\begin{appendices}




\section{Radial differential equation}\label{A1}
The radial component of transformed Klein-Gordon equation can be expressed as
\begin{equation}
	z(1-z)\frac{d^{2}R(z)}{dz^2}+(1-z)\frac{dR(z)}{dz}+\left[\frac{A}{z}+B+\frac{C}{1-z}\right]R(z)=0.
\end{equation}
The ingoing solution of the above differential equation at the event horizon is
\begin{equation}
	R_{in}(z)=k_{1}z^{-i\sqrt{A}}\left(1-z\right)^{L+1}{}_2 F_1\left(a_s,b_s;c_s;z\right),
\end{equation}
where $a_s$,$b_s$, and $c_s$ are
\begin{equation}
	\begin{split}
		a_s&=1+L-i\left(\sqrt{A}+\sqrt{-B}\right),\hspace{3mm}
		b_s=1+L-i\left(\sqrt{A}-\sqrt{-B}\right)\\
		c_s&=1-2i\sqrt{A}.
	\end{split}
\end{equation}
\section{Hypergeometric relations for large $r$}\label{A2}
\begin{equation}
	\begin{split}
		&{}_2 F_1\left(a,b;c;z\right)=\frac{\Gamma(c)\Gamma(c-a-b)}{\Gamma(c-a)\Gamma(c-b)}{}_2 F_1\left(a,b,a+b-c+1;1-z\right)\\&+\left(1-z\right)^{c-a-b}
		\frac{\Gamma(c)\Gamma(a+b-c)}{\Gamma(a)\Gamma(b)}
		{}_2 F_1\left(c-a,c-b;c-a-b+1;1-z\right)
	\end{split}	
\end{equation}

\section{Relation between imaginary gamma functions}\label{A3}
\begin{equation}
	\begin{split}
		&\frac{\Gamma\left(1+L-2i\frac{\omega r_{+}^2}{k\left(r_{+}-r_{-}\right)}\right)}
		{\Gamma\left(-L-2i\frac{\omega r_{+}^2}{k\left(r_{+}-r_{-}\right)}\right)}
		=\frac{\Gamma\left(1+L-2i\frac{\omega r_{+}^2}{k\left(r_{+}-r_{-}\right)}\right)}
		{\Gamma\left(-L-2i\frac{\omega r_{+}^2}{k\left(r_{+}-r_{-}\right)}\right)}
		\frac{\Gamma\left(1+L+2i\frac{\omega r_{+}^2}{k\left(r_{+}-r_{-}\right)}\right)}
		{\Gamma\left(1+L+2i\frac{\omega r_{+}^2}{k\left(r_{+}-r_{-}\right)}\right)}\\
		&=\frac{\left|\Gamma\left(1+L-2i\frac{\omega r_{+}^2}{k\left(r_{+}-r_{-}\right)}\right)\right|^2}{\pi}
		\sin\left[\pi\left(1+L+2i\frac{\omega r_{+}^2}{k\left(r_{+}-r_{-}\right)}\right)\right]\\
		&=\frac{\left|\Gamma\left(1+L-2i\frac{\omega r_{+}^2}{k\left(r_{+}-r_{-}\right)}\right)\right|^2}{\pi}
		\left[-\sin\left(\pi L\right)
		\cosh\left(2\pi\frac{\omega r_{+}^2}{k\left(r_{+}-r_{-}\right)}\right)
		\right.\\ &\left.-i\cos\left(\pi L\right)\sinh\left(2\pi\frac{\omega r_{+}^2}{k\left(r_{+}-r_{-}\right)}\right)
		\right]
	\end{split}
\end{equation}
\end{appendices}


\bibliography{sn-bibliography}

\end{document}